\documentstyle[12pt]{article}
\def\fun#1#2{\lower3.6pt\vbox{\baselineskip0pt\lineskip.9pt
\ialign{$\mathsurround=0pt#1\hfil##\hfil$\crcr#2\crcr\sim\crcr}}}

\newcommand{\be}{\begin{equation}}
\newcommand{\ee}{\end{equation}}
\newcommand{\bd}{\begin{displaymath}}
\newcommand{\ed}{\end{displaymath}}
\newcommand{\ba}{\begin{array}}
\newcommand{\ea}{\end{array}}
\newcommand{\bt}{\begin{tabular}}
\newcommand{\et}{\end{tabular}}
\newcommand{\bc}{\begin{center}}
\newcommand{\ec}{\end{center}}

\begin{document}

\titlepage

\hfill {\large\bf E2--98--236}

\hfill {\large\bf August 1998}

\begin{center}
{\large $\bf D_s$-- mesons inclusive hadroproduction \\
in the Quark -- Gluon String Model}\\[0.5cm]
G.H. Arakelyan \footnote{Yerevan Physics Institute,
 Yerevan 375036, Armenia \\E--mail:argev@vxitep.itep.ru~~~~
argev@nusun.jinr.ru} \\[0.5cm]
Joint Institute for Nuclear Research, Dubna, Russia \\[0.5cm]
Sh.S. Yeremyan \\[0.5cm]
Yerevan Physics Institute, Yerevan 375036, Armenia \\[0.5cm]
\end{center}


\vspace{1.5cm}

\begin{abstract}

The hadroproduction of $D_s$ mesons is discussed within the
framework of the modified Quark-Gluon String model (QGSM) taking
into account the decays of corresponding $S$--wave resonances.
A description of the existing experimental data on inclusive
spectra  of $D_s$ - mesons production in $\pi p$ - collision and
asymmetry of leading ($D_s^-$) and nonleading ($D_s^+$) mesons in
$\Sigma^- p$-- interaction is obtained.
The predictions for cross sections and leading/nonleading asymmetry
in $\pi p$ - and $\Sigma^- p$-- collision for different initial
momenta are also given.

\end{abstract}

\vfill

\noindent

\newpage


~~~~~Hadroproduction of charmed particles is now being investigated
both experimentally and theoretically.
On the one hand the methods, available for light hadrons such as quark
models, symmetries and so on, can be applied to hadrons, containing
the charmed quark. On the other hand, the presence of $c$- quark inside
the charmed hadrons allows one to use QCD methods.

There is wide variety of theoretical models which are more or less
satisfactory applied to description of charmed particle hadroproduction
(see \cite{SHSURV} and references therein).

Among these models the Quark--Gluon String model (QGSM) was successfully
used for the description of many features of multiparticle production
in hadron-hadron collisions both for light and charmed hadrons.

The modification of the QGSM taking into account the
contributions from decays of corresponding $S$--wave resonances
was developed in \cite{AVZA}.

In the previous papers \cite{AVZA,AVC} this approach was used
for the calculation of  charmed hadron (meson and baryon) spectra,
produced by different hadron beams ($\pi$, $p$, $\Sigma$ and $\Xi$)
without taking into account the charmed sea contribution.

The charmed sea contribution in QGSM for the $D$ mesons hadroproduction
were considered in \cite{AD,PY60} to describe the experimental data of
leading/nonleading $D$-- mesons asymmetry.

The predictions for inclusive spectra of $D_s$-- mesons
produced by different hadron beams in the various modifications of QGSM
without taking into account the contribution of charmed sea were
calculated earlier in \cite{AVZA,PDS}. However in these papers,
due to the absence of the measurements, there was not made the
comparison of the model calculations with experimental data.

In this note we analyze for the first time  the recently appeared
experimental data \cite{WA92PPE} - \cite{AIFP97} on inclusive spectra
and production asymmetry of the leading/nonleading charmed-strange
$D_s^- (s \bar c)$ -- and $D_s^+ (c \bar s)$-- pseu\-do\-sca\-lar
mesons. The analysis has been performed in the framework of modified
QGSM \cite{AVZA} taking into account the contribution of vector
$D_s^*$ resonance decay into
$D_s$ through $D_s^* \rightarrow D_s \gamma$ mode \cite{RPP}.

 The formulae for inclusive spectra of $D_s^*$ mesons hadroproduction
for different beams and the expressions for quark and diquark
distribution functions in the initial hadrons were given  in
\cite{AVZA,AD}.

  Here we used the parametrization of charmed sea in $\pi^-$ and
$\Sigma^-$ given in \cite{AD}.



In the present paper we use the following
parametrization of the quark and diquark fragmentation functions into
$D_s$- mesons:

\be
\label{13}
G_{d}^{D_s^{+}}(z)=(1-z)^{\lambda-\alpha_{\psi}(0)+
2(1-\alpha_{R}(0))+2(\alpha_R(0)-\alpha_{\varphi}(0))},
\ee

\begin{eqnarray}
\label{133}
G_{d}^{D_s^{+}}(z)=G_{d}^{D_s^{-}}(z)=G_{u}^{D_s^{-}}(z)=
G_{u}^{D_s^{-}}(z)=G_{\bar d}^{D_s^{-}}(z)= \nonumber \\
G_{\bar d}^{D_s^{-}}(z)=G_{\bar u}^{D_s^{-}}(z)=G_{\bar u}^{D_s^{-}}(z)
\end{eqnarray}

\be
\label{14}
{G}_{s}^{D_s^{-}}(z)=(1-z)^{\lambda-\alpha_{\psi}(0)}
(1+a_{1}^{D}z^{2})
\ee

\be
\label{15}
{G}_{s}^{D_s^{+}}(z)=(1-z)^{\lambda-\alpha_{\psi}(0)+
2(1-\alpha_{\varphi}(0))}
\ee

\begin{equation}
\label{16}
G_{c(\bar{c})}^{D_s^{-}(D_s^{+})}(z)=\frac{b^{D_s}}{a^{D_s}_0}
z^{1-\alpha_{\psi}(0)}(1-z)^{\lambda-\alpha_{\varphi}(0)}.
\end{equation}

\be
\label{17}
{G}_{dd}^{D_s^+}(z)=(1-z)^{\lambda-\alpha_{\psi}(0)+2(1-\alpha_N(0))
+(\alpha_R(0)-\alpha_{\varphi}(0))}
\ee

\be
\label{18}
{G}_{dd}^{D_s^-}(z)={G}_{dd}^{D_s^+}(z)={G}_{uu}^{D_s^-}(z)=
{G}_{uu}^{D_s^+}(z)
\ee

\be
\label{19}
{G}_{ds}^{D_s^-}(z)=(1-z)^{\lambda+\alpha_R(0)-2\alpha_N(0)
+\alpha_R(0)-\alpha_{\psi}}(\frac{1+a_1z^2}{2}+\frac{(1-z)^2}{2})
\ee

\be
\label{20}
{G}_{ds}^{D_s^+}(z)=(1-z)^{\lambda-\alpha_{\psi}(0)+2(1-\alpha_N(0))
+2(\alpha_R(0)-\alpha_{\varphi}(0))}
\ee

where $\lambda=0.5$, $\alpha_R(0)=0.5$, $\alpha_N(0)=-0.5$. Under
these calculations we consider two values of the intercept of charmed
$c \bar c$-- trajectory $\alpha_{\psi}(0)=-2.18$ and
$\alpha_{\psi}(0)=0.$.


Now we turn to the comparison of the existing experimental data  
\cite{WA92PPE} - \cite{AIFP97} for $D_s$--mesons produced in 
$\pi^-p$ and $\Sigma^- p$ collisions with the QGSM calculations 
in the framework of the model under consideration using 
fragmentation functions (\ref{13}) - (\ref{20}).
The values of free parameters were obtained by fitting to the 
experimental data.

In what follows, we will present four curves in all figures.
They correspond to two values of $\alpha_{\psi}$ trajectory and
presence or absence of the charmed sea contribution:

1) $\alpha_{\psi}=-2.18$ (full line) with corresponding values of
parameters  $a_0^D=0.0007$, $a_1^D=5$,  $b^{D_s}=1.6$,
$\delta_c=0.05$,

2) $\alpha_{\psi}=0$ (dashed line),$a_0=0.0005 $, $a_1=5$,
$b^{D_s}=1.6$,$\delta_c=0.05$,

3) $\alpha_{\psi}=-2.18$ (dashed-dotted line) - the same as in 1)
without charmed sea contribution $\delta_c=0$ and

4) $\alpha_{\psi}=0$ (dotted line) -  the same as in 2) but
$\delta_c=0$.

The charmed sea suppression factor $\delta_c=0.005$ was obtained
earlier in  \cite{AD}.

All theoretical curves in the model under consideration are
sums of the directly produced $D_s$--meson cross section and the
contribution from the decay of the corresponding
$D^*_s \rightarrow D \gamma$ resonance.

The experimental data on the $x_F$-- dependence of the differential
cross section of $D_s^+$--meson in  $\pi^-p$-- interaction
at initial momenta $P_L=230\;GeV$ \cite{BZC49} together with model
calculations are presented in Fig.1. Note that the normalization of
the experimental data on this figure is arbitrary and here we may
compare only the shape of the cross section.

The model calculations for the spectra of the sum of all $D_s$--
mesons in the reaction $\pi^- p \rightarrow D_s X$ at $P_L=350\;GeV/c$
are compared with the experimental data at $P_L=350\;GeV/c$
\cite{WA92PPE} in Fig.2. It is necessary to mention that in  the
$\pi p$-- collision the production cross sections of
$D_s^- (s \bar c)$ and $D_s^+ (c \bar s)$-- mesons are equal because
they do not contain the valence quarks from pion beam. The experimental
measurements of the $D_s^-$ ¨ $D_s^+$-- mesons total cross sections ,
given in \cite{WA92PPE}, slightly differ each other, although they are
close within the rather large experimental errors.

In Fig.3 the experimental data on the total cross section
versus momentum of the initial $\pi$ meson in $x_F>0$ region are 
compared with the model calculations. As we can see the existing 
experimental data do not allow to draw a definite conclusion on 
momentum dependence or, as it was noted in \cite{WA92PPE}, there is no  
indication of momentum dependence which contradicts to PYTHIA (see 
curve on Fig.9c in \cite{WA92PPE}) and our calculations. It is also 
necessary to note that the first point at $P_L=235\;GeV/c$ (NA32) 
\cite{BZC49} stands for the $D_s^+$ meson cross section, while two 
others - (E769) at $P_L=250\;GeV/c$ \cite{ALPRL77F} and (WA92) at 
$P_L=350\;GeV/c$ \cite{WA92PPE} were given for the sum of $D_s^+$ 
and $D_s^-$ mesons.

The predictions of the model for the inclusive spectra of $D_s^{+}$
 mesons in $\pi^- p$-- collision at $P_L=500\;GeV/c$ and $\Sigma^-p$
collisions at  $P_L=330\;GeV/c$ and $600\;GeV/c$ are given in 
Figs.4 - 8.

The $x_{F}$ dependence of leading/nonleading production asymmetry
€($D_s^-$,$D_s^+$) for $\Sigma^- p$-- collision together with
experimental data of the WA89 collaboration for
$P_L=340\;GeV/c$ \cite{WA89AS} is presented in Fig.9.
The predictions for the same asymmetry at $P_L=600\;Ē'/c$ is shown 
in Fig.10.

We do not present the comparison of our calculations for
 asymmetry €($D_s^-$,$D_s^+$) in $\pi^- p$ collision with the
experimental measurements \cite {AIFP97} because it is equal to zero
due to equality of  $D_s^-$ and $D_s^+$ mesons cross sections.

As we can see from the comparison with the experimental data, the
model under consideration represents the shape of $D_s$-- mesons cross
section in  $\pi^- p$-- collision. Slightly better agreement is for
the case of charmed trajectory intercept equal to  $\alpha_{\psi}=0$.
Recent experimental measurements of leading/nonleading asymmetry
€($D_s^-$,$D_s^+$) in the $\Sigma^- p$-- interaction at 
$P_L=340\;Ē'/c$ \cite{WA92PPE} prefer the $\alpha_{\psi}=-2.18$ 
without charmed sea contribution (curve 3 on fig.9). Unlike the  
$\pi$ induced reactions where, as it was already noted, the asymmetry 
is equal to zero, in the hyperon induced reaction it should be 
significant due to the presence of the valence $s$-- quark which was 
testified by WA89 collaboration measurements \cite{WA89AS}. However 
in the description of inclusive cross section the variant 3) gives 
the rapidly falling cross section.
Taking into account the charmed sea contribution leads to
decreasing of the cross section, especially at large  $x_F$, but
leads to de decreasing of the asymmetry at the same region.
Unfortunately existing experimental data due to absence of the
cross sections measurements at $x_F>0.5$ do not allow to do the
unambiguous choice between considered variants.

In the model under consideration it is possible to improve the
description of the experimental data \cite{WA92PPE} by slightly 
changing the value of the $a_1$ parameter in the fragmentation 
functions of the $s$-- quark (\ref{14}) and $ds$-- diquark 
(\ref{19}).
However for the exact determination of the quark and diquark
fragmentation functions into the  $D_s$-- mesons  more
precise measurements of all observations are needed: total and 
differential cross sections and leading/nonleading production 
asymmetry especially at large $x_F$.

{\it Acknowledgments}. We are grateful to A.B.Kaidalov, K.G.Bo\-res\-kov
and A.Capella for useful discussions. The work of A.G. was partly
supported by INTAS grant 93-0079ext, NATO grant OUTR.LG 971390
and RFBR grant 98-02-17463.

\newpage

\section*{Figure captions}

\noindent

Fig.1. Comparison of the QGSM calculations with the experimental data
(NA32) \cite{BZC49} on differential cross sections of the $D_s^+$--meson
production in $\pi^-p$-- interaction at $P_L=230\;GeV/c$. The
theoretical curves notification: 1) full curve were calculated for
$\alpha_{\psi}=-2.18$, 2) dashed curve stands for $\alpha_{\psi}=0$
(in 1 and 2 with the contribution of the charmed sea), 3)
dashed - dotted line - $\alpha_{\psi}=-2.18$, 4) dotted line -
$\alpha_{\psi}=0$ ( in 3 and 4 the charmed sea contribution is absent).
\\[2mm]

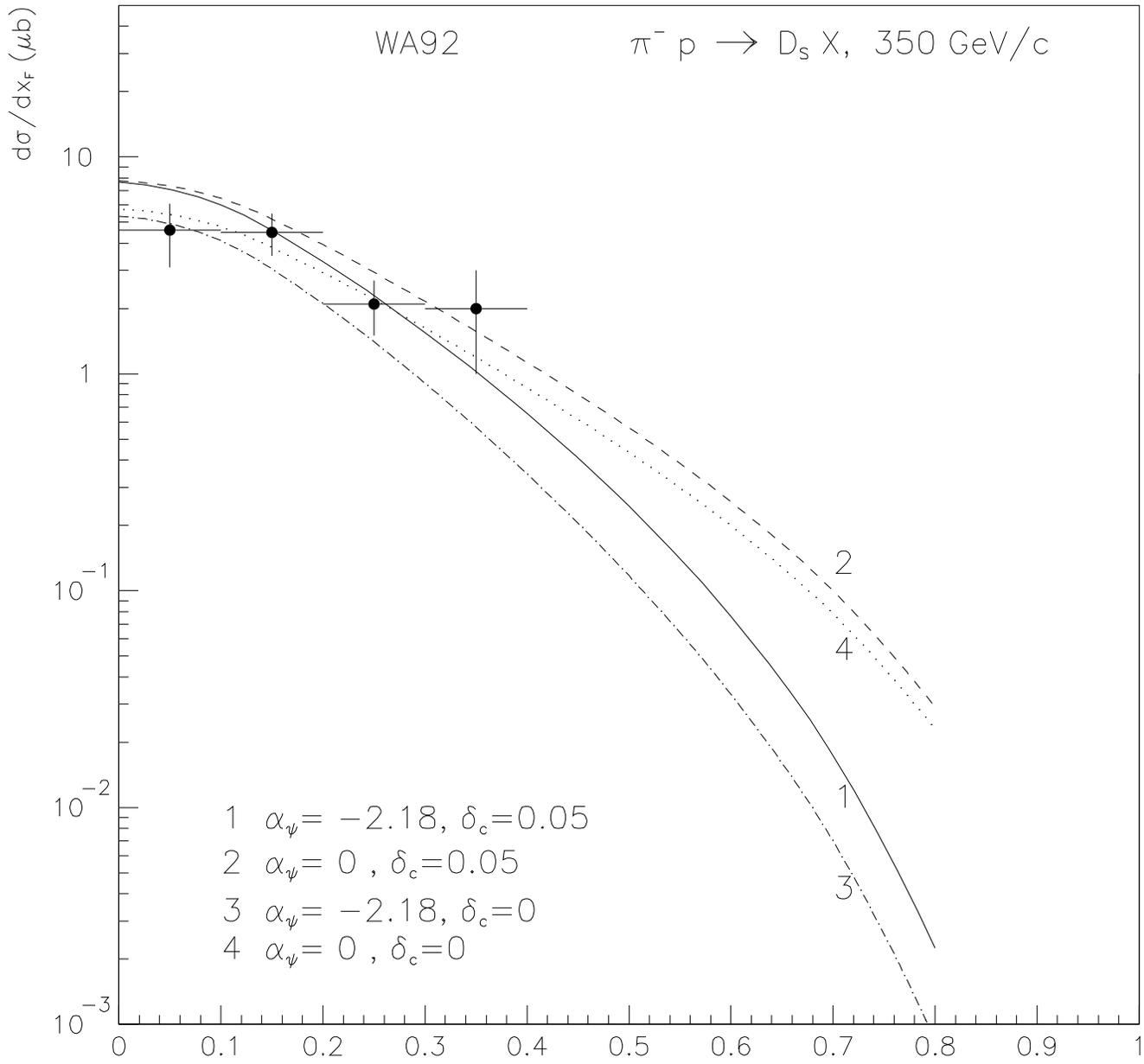
Fig.2. Comparison of the QGSM calculations with the experimental data
(WA92) \cite{WA92PPE} of the summary spectra  of $D_s$--mesons
production in $\pi^-p$-- interaction at  $350\;GeV/c$. The curves are
the same as in Fig.1. \\[2mm]

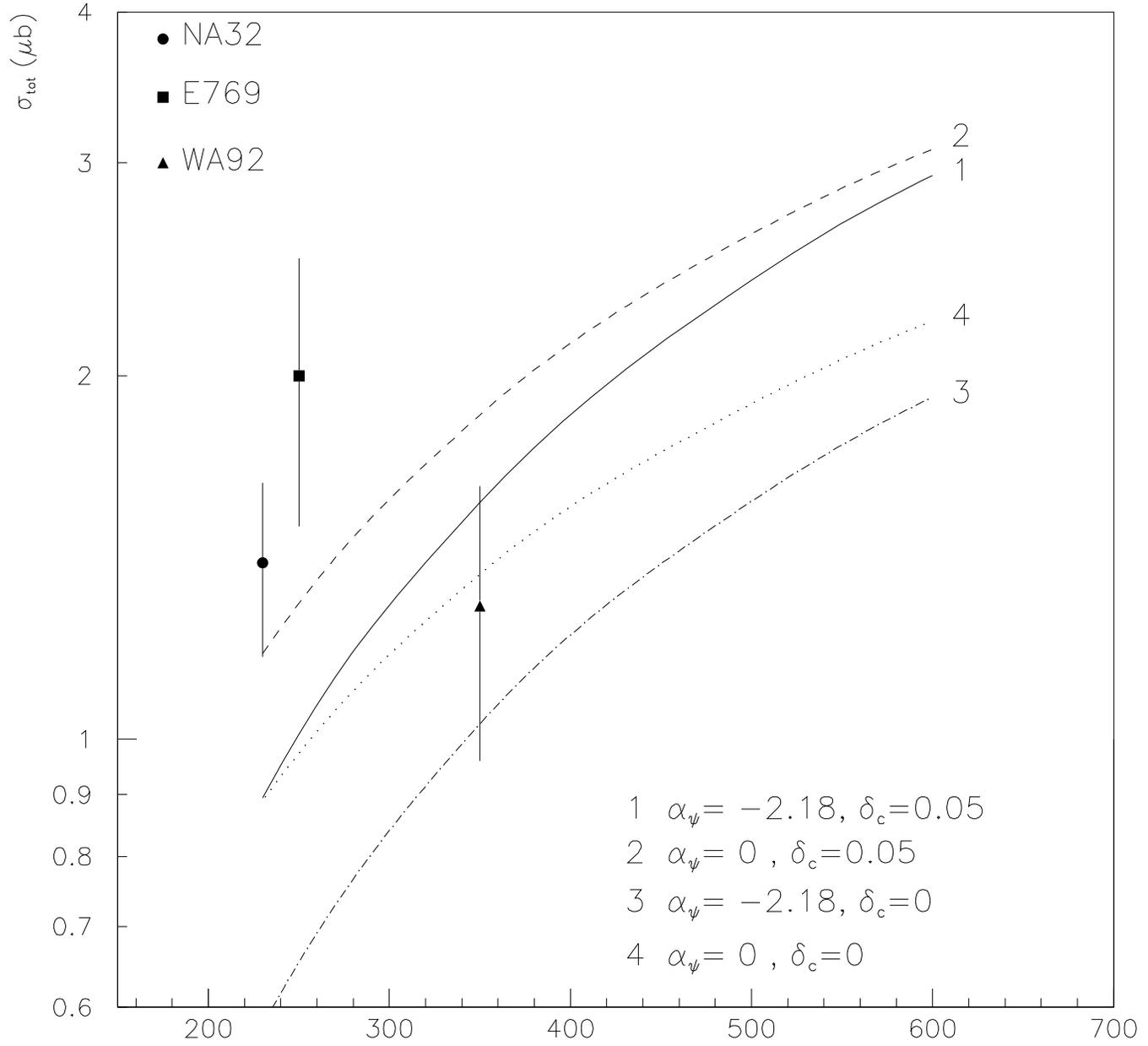
Fig.3. Comparison of the QGSM calculations with the experimental
data on $D_s$ mesons total cross sections in $\pi^-p$-- interaction.
The curves are the same as in Fig.1. \\[2mm]

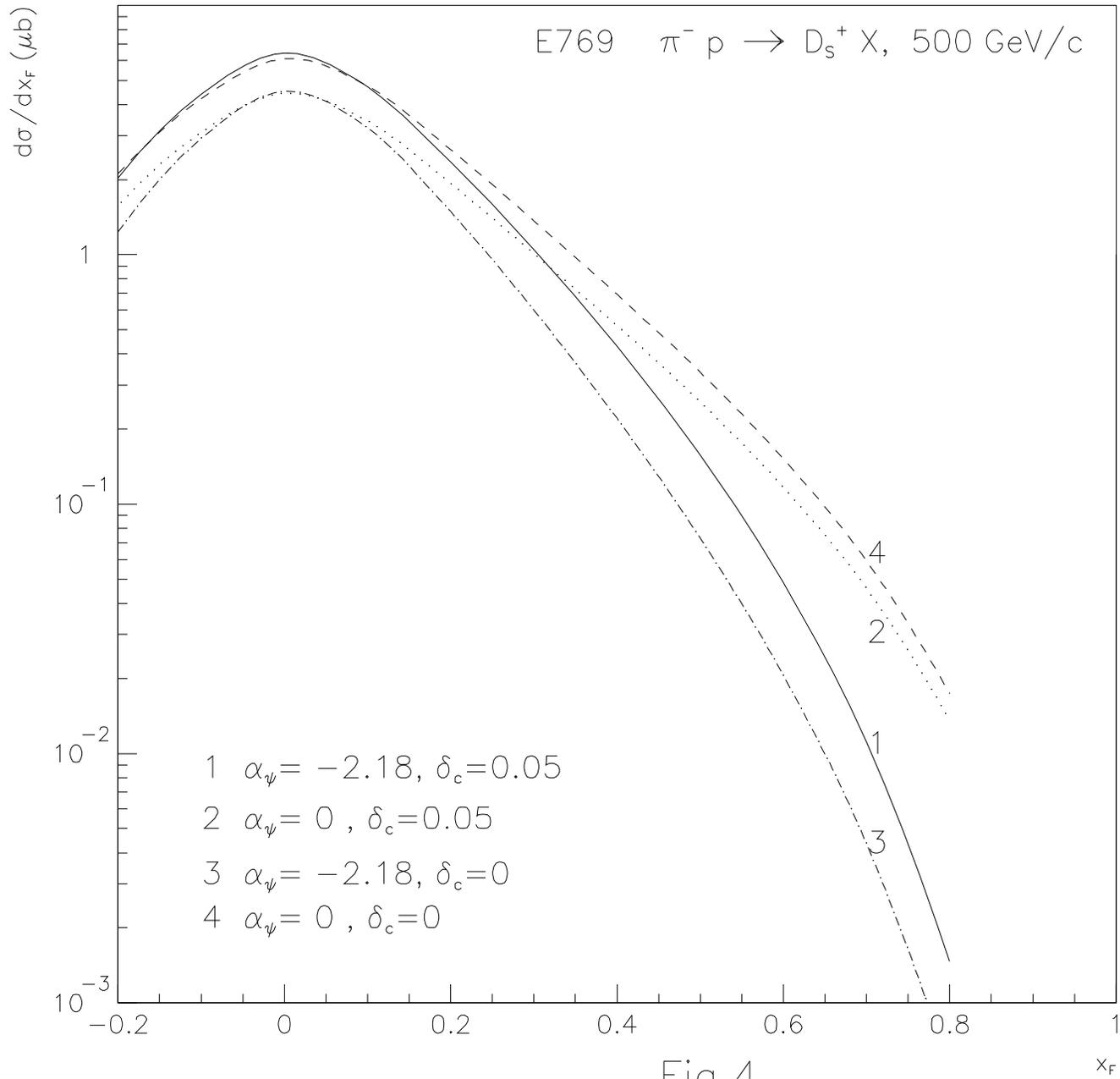
Fig.4. Predictions for the differential cross sections of the
$x_F$--dependence of the $D_s^+$--meson production in $\pi^- p$--
interaction at $P_L=500\;GeV/c$ (E769). The curves are the same
as in Fig.1. \\[2mm]

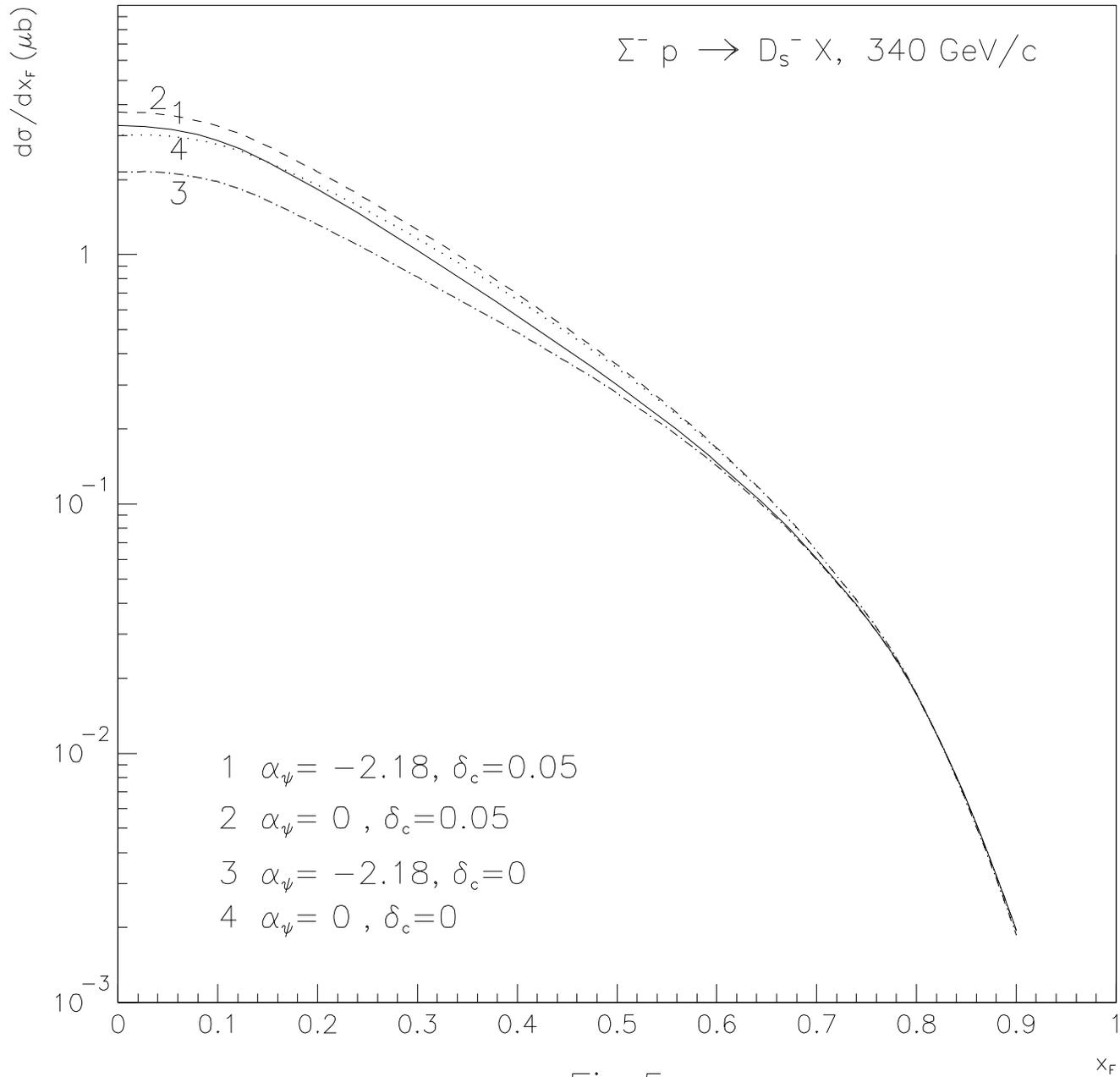
Fig.5. Model prediction for the $x_F$--dependence of the
$D^{-}_s$--meson cross section in $\Sigma^-p$ interaction at
$340\;GeV/c$. The curves are the same as in Fig.1. \\[2mm]

Fig.6. Model prediction for the $x_F$--dependence of the
$D^{+}_s$--meson cross section in $\Sigma^-p$ interaction at
$340\;GeV/c$. The curves are the same as in Fig.1. \\[2mm]

Fig.7. Model prediction for the $x_F$-- dependence of the
$D^{-}_s$--meson cross section in $\Sigma^-p$ interaction at
$600\;GeV/c$. The curves are the same as in Fig.1. \\[2mm]

Fig.8. Model prediction for the $x_F$--dependence of the
$D^{+}_s$--meson cross section in $\Sigma^-p$ interaction at
$600\;GeV/c$. The curves are the same as in Fig.1. \\[2mm]

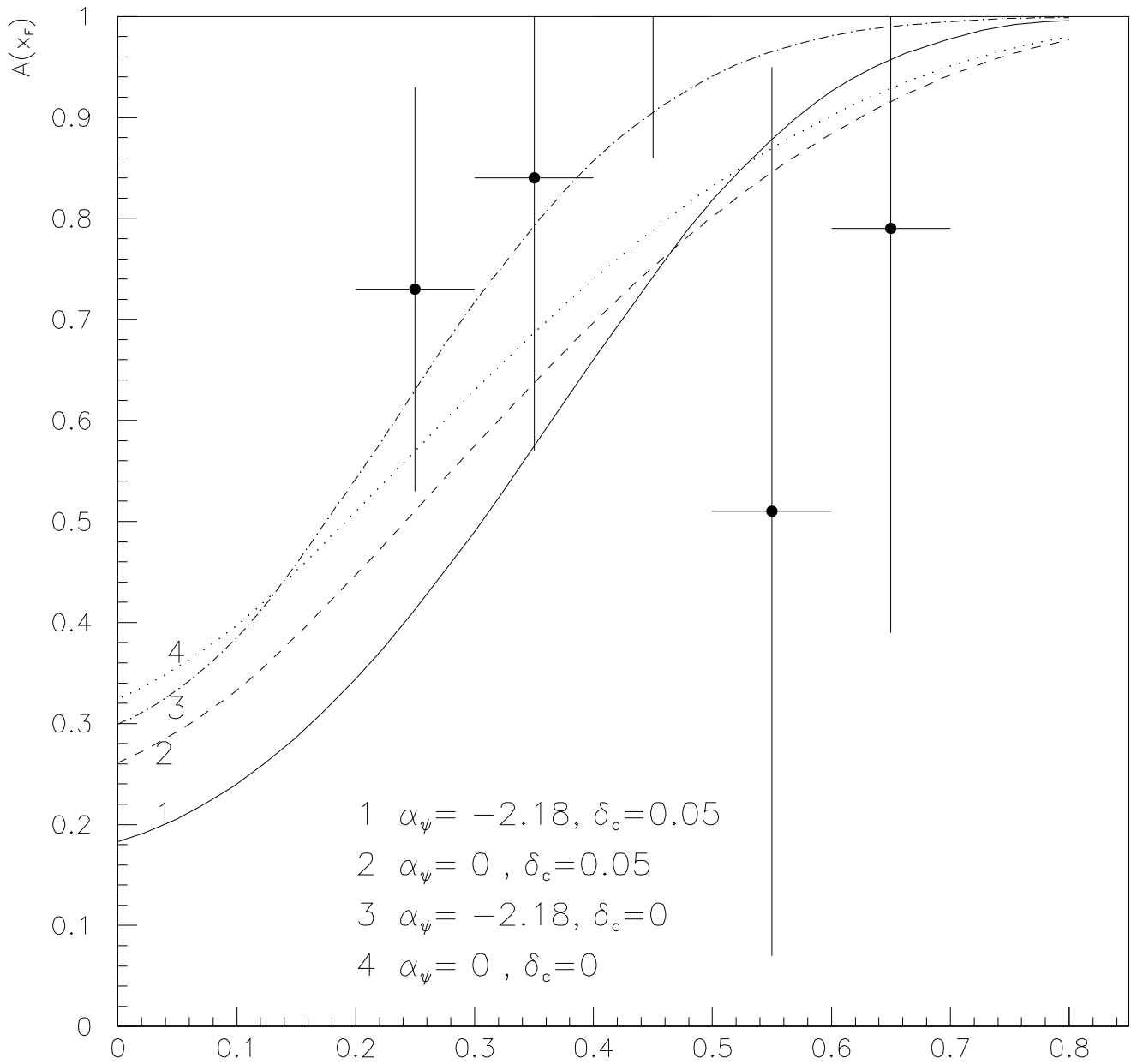
Fig.9.Comparision of the model calculations for the $x_F$--dependence
of $D_s^-/D_s^+$ asymmetry for the  $\Sigma^-$ beam at $340\;GeV/c$
with the experimental data \cite{WA89AS}.
The curves are the same as in Fig.1. \\[2mm]

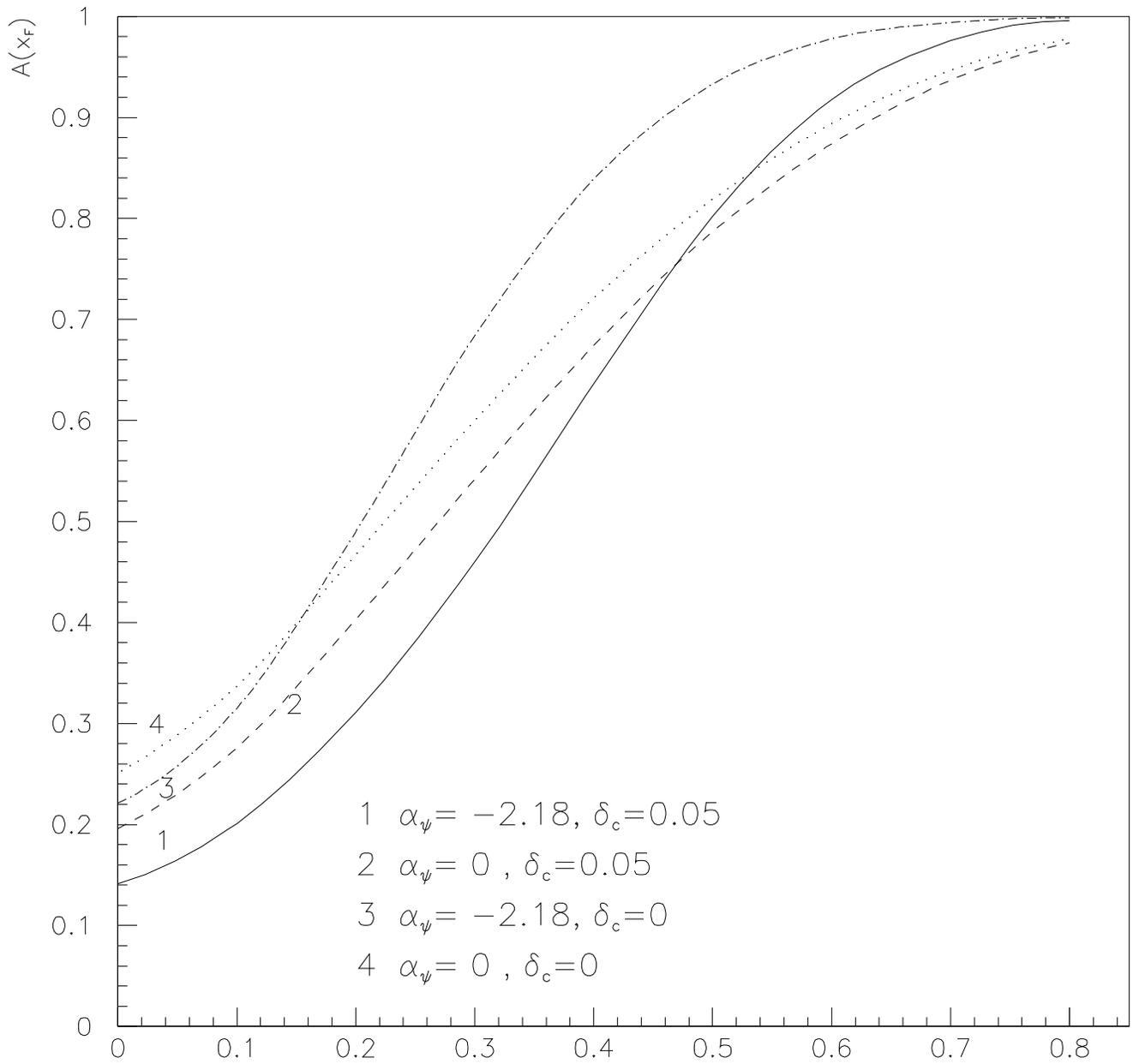
Fig.10. Predictions for the $x_F$--dependence of $D_s^-/D_s^+$
asymmetry for the  $\Sigma^-$ beam at $600\;GeV/c$.
The curves are the same as in Fig.1. \\[2mm]


\end{document}